\documentclass[secthm]{elsart}
\usepackage{amssymb}

\begin{document}
%\sloppy

\begin{frontmatter}

\title{Quantum Scholasticism: On Quantum Contexts,  Counterfactuals, and the Absurdities of Quantum Omniscience}

\author{Karl Svozil}
\address{Institut f\"ur Theoretische Physik, University of Technology Vienna,
Wiedner Hauptstra\ss e 8-10/136, A-1040 Vienna, Austria}
\ead{svozil@tuwien.ac.at}
\ead[url]{http://tph.tuwien.ac.at/$\widetilde{\;\;}$svozil/}

\begin{abstract}
Unlike classical information, quantum knowledge is restricted to the outcome of measurements of maximal observables corresponding to single contexts.
\end{abstract}

\begin{keyword}
maximal observables, context, quantum information, omniscience

\PACS 01.70.+w, 01.65.+g, 03.67.-a,02.50.-r
\end{keyword}

\end{frontmatter}

%...if some careful and meditative mind were to take
%the trouble to clarify and direct their thoughts in the manner of analytic geometers,
%he would find a great treasure of very important truths, wholly demonstrable.

\section{Introduction}

The violation of classical bounds
\cite{Boole,Boole-62}
on joint quantum probabilities
enumerated by Bell \cite{bell,pitowsky},
Clauser-Horn-Shimony-Holt (CHSH) \cite{chsh,clauser}
and others
\cite{peres222,2000-poly,collins-gisin-2003,filipp-svo-04-qpoly-prl},
the Kochen-Specker (KS)
\cite{specker-60,kochen1,ZirlSchl-65,Alda,Alda2,kamber64,kamber65,svozil-tkadlec,cabello-96}
as well as the Greenberger-Horn-Zeilinger (GHZ)
\cite{ghz,ghsz,mermin-93}
theorems provide constructive, finite proofs that
observables outside of a single quantum context
cannot consistently co-exist.
Here, the term {\em context}
refers to a maximal collection of co-measurable observables associated with commuting operators.
Every context can also be characterized by a single (but nonunique) maximal operator.
All operators within a context are functions thereof
(see Ref.~\cite{v-neumann-49}, Sec.~II.10, p. 90, English translation p.~173 and Ref.~\cite{halmos-vs}, Sec.~84).
In quantum logic
\cite{birkhoff-36,ma-57,pulmannova-91,svozil-ql},
contexts are represented by Boolean subalgebras or blocks
pasted together to form the Hilbert lattice.
(For the sake of nontriviality, Hilbert spaces of dimension higher than two are considered.)
In an algebraic sense, a context represents a ``classical mini--universe,''
which is distributive and allows for as many two--valued states --- interpretable as classical truth assignments ---
as there are atoms.

By definition, no direct measurement of observables ``outside'' of a single context is possible.
Therefore, any assumption about the physical existence of such observables
results in the invocation of counterfactuals.
For example, Einstein, Podolsky and Rosen (EPR) \cite{epr}
suggested to measure and counterfactually infer two contexts simultaneously
by considering elements of physical reality which cannot be measured simultaneously
on the same quantum.
In this respect, quantum physics relates to scholastic philosophy.
Indeed, in an informal paper \cite{specker-60} announcing KS, Specker  explicitly referred
to  the ``infuturabilities'' of scholastic philosophy.

Related to counterfactuals is the idea of a (divine) {\em omniscience} ``knowing'' all the
factuals and counterfactuals in the naive sense that
``if a proposition is true, then an omniscient agent (such as God) knows that it is true.''
Already Thomas Aquinas
considered questions such as whether God has knowledge of non-existing things
(Ref.~\cite{Aquinas}, Part 1, Question 14, Article 9) or things that are not yet
(Ref.~\cite{Aquinas}, Part 1, Question 14, Article 13).

In classical physics, there is just one global context which is
trivially constituted by all conceivable observables.
Hence, there is no conceptual or principal reason to assume counterfactuals;
sometimes they are just considered for convenience (saving the experimenter from measuring redundant observables).
The empirical sciences implement classical omniscience by assuming that
in principle all observables of classical physics are (co-)\-measurable without any restrictions.
No distinction is made between an observable obtained
by an ``actual'' and a ``potential'' measurement.
Precision and (co-)\-measurability are limited only by the technical capacities of the experimenter.
The principle of empirical classical omniscience has given rise to the realistic believe that
all observables ``exist,'' regardless of the state preparations and observations.
Physical (co-)\-existence is thereby related to the realistic assumption \cite{stace}
(sometimes referred to as the ``ontic'' \cite{atman:05} viewpoint) that such physical entities exist
even without being experienced by any finite mind.

Formally, counterfactuals and classical omniscience are supported by the following two properties.
\begin{itemize}
%(i)
\item[(i)]  Boolean logics and absence of complementarity:
Historically, the discovery of the uncertainty principle and quantum complementarity
marked a first departure from classical omniscience.
A formal expression of complementarity is the nondistributive algebra of quantum observables.
Alas,
nondistributivity of the empirical logical structure is no sufficient
condition for the impossibility of omniscience.
For example, both generalized urn models \cite{wright:pent,wright}
as well as equivalent \cite{svozil-2001-eua} finite automata
\cite{e-f-moore,schaller-96,dvur-pul-svo,cal-sv-yu}
exhibit complementarity, yet they
possess two--valued states
interpretable as omniscience; i.e.,
as global truth assignments with a consistent value
``0'' ({\em false})
or
``1'' ({\em true})
for every observable.
%(ii)
\item[(ii)]
``Abundance'' of two--valued states interpretable as omniscience of the system:
Thereby, any such ``dispersionless'' two--valued state --- associated with a classical ``truth table''
--- can be defined on all observables,
regardless of whether they have been actually observed or not.
\end{itemize}

In contrast,
quantum propositions neither satisfy distributivity,
nor do they support two--valued states.
Recall
Schr\"odinger's interpretation of the quantum wave function
(in \S  7 of Ref.~\cite{schrodinger}) in terms of a
``catalogue of expectations.''
Every page of this catalogue of expectations is represented by a single context.
In quantum mechanics,
(as well as in quasi-classical models \cite{svozil-2001-eua}),
due to complementarity, contexts are not global,
and the structure of contexts as well as the probability measures on them
\cite{Gleason,r:dvur-93} pose many
challenging questions.

\section{``Scarcity'' of two--valued states}

Gleason's theorem \cite{Gleason,r:dvur-93} states that the quantum probabilities
can be derived from the assumption
that classical probability theory holds within contexts.
Yet, unlike classical systems,
they are no convex combination of global
two--valued states.
Formally, this is due to the fact that the quantum propositions do not support
globally defined two--valued states.

What happens if one insist in the use of two--valued states
outside of a single context
by considering quantum propositional structures
still allowing ``a few'' two--valued states?
In this case,
the invocation of counterfactuals and
the ``scarcity'' of two--valued states accounts for some consequences
which, depending on the disposition of the recipient, appear ``mindboggling'' to absurd.

By bundling together propositional structures
giving rise to such ``mindboggling''
properties, one arrives at the KS conclusion.
For such  finite compositions of observables,
the mere assumption of a globally defined truth table results
in a complete contradiction.
Alas, by contemplating the situation not bottom--up as usually, but top--down; i.e., from the point of view of KS,
it is not too difficult to derive ``mindboggling'' statements from absurdities.
Indeed, the  {\em principle of explosion} (stating that {\it ex falso quodlibet,} or {\it contradictione sequitur quodlibet})
which, due to the pasting construction of Hilbert lattices, holds also in quantum logic,
implies that ``anything follows from a contradiction.''

\subsection{Dual Greechie and Tkadlec diagrams}

For a proof of the ``scarcity'' of two--valued states,
Greechie diagrams symbolizing one-dimensional projectors by points
and contexts by maximal smooth unbroken curves are considered.
The ``dual'' Tkadlec diagrams \cite{tkadlec-00}
symbolize entire contexts by points,
and links between contexts by lines joining them.

\begin{figure}%[p]
\begin{center}
\begin{tabular}{ccc}
%TeXCAD Picture [1.pic]. Options:
%\grade{\on}
%\emlines{\off}
%\epic{\off}
%\beziermacro{\on}
%\reduce{\on}
%\snapping{\off}
%\pvinsert{% Your \input, \def, etc. here}
%\quality{8.000}
%\graddiff{0.005}
%\snapasp{1}
%\zoom{8.0000}
\unitlength 0.2mm % = 2.845pt
\linethickness{0.8pt}
\ifx\plotpoint\undefined\newsavebox{\plotpoint}\fi % GNUPLOT compatibility
\begin{picture}(151.75,61.75)(0,0)
\put(150,0){\circle{7}}
\put(100,0){\circle{7}}
\put(50,0){\circle{7}}
\put(0,0){\circle{7}}
\put(150,20){\circle{7}}
\put(100,20){\circle{7}}
\put(50,20){\circle{7}}
\put(0,20){\circle{7}}
\put(150,40){\circle{7}}
\put(100,40){\circle{7}}
\put(50,40){\circle{7}}
\put(0,40){\circle{7}}
\put(150,60){\circle{7}}
\put(100,60){\circle{7}}
\put(50,60){\circle{7}}
\put(0,60){\circle{7}}
\put(150,0){\line(0,1){60}}
\put(100,0){\line(0,1){60}}
\put(50,0){\line(0,1){60}}
\put(0,0){\line(0,1){60}}
\put(0,0){\line(1,0){150}}
\put(75,5){\makebox(0,0)[cc]{$a$}}
\put(10,50){\makebox(0,0)[cc]{$b$}}
\put(60,50){\makebox(0,0)[cc]{$c$}}
\put(110,50){\makebox(0,0)[cc]{$d$}}
\put(160,50){\makebox(0,0)[cc]{$e$}}
\end{picture}
&
$\qquad$
$\qquad$
&
%TeXCAD Picture [1.pic]. Options:
%\grade{\on}
%\emlines{\off}
%\epic{\off}
%\beziermacro{\on}
%\reduce{\on}
%\snapping{\off}
%\pvinsert{% Your \input, \def, etc. here}
%\quality{8.000}
%\graddiff{0.005}
%\snapasp{1}
%\zoom{8.0000}
\unitlength 0.125mm % = 2.845pt
\linethickness{0.8pt}
\ifx\plotpoint\undefined\newsavebox{\plotpoint}\fi % GNUPLOT compatibility
\begin{picture}(103.5,103.75)(0,0)
\put(100,100){\line(-1,-1){100}}
\put(0,100){\line(1,-1){100}}
\put(100,100){\circle*{15}}
\put(100,0)  {\circle*{15}}
\put(50,50)  {\circle*{15}}
\put(0,0)    {\circle*{15}}
\put(0,100)  {\circle*{15}}
\put(70,50){\makebox(0,0)[cc]{$a$}}
\put(120,100){\makebox(0,0)[cc]{$b$}}
\put(120,0){\makebox(0,0)[cc]{$c$}}
\put(20,0){\makebox(0,0)[cc]{$d$}}
\put(20,100){\makebox(0,0)[cc]{$e$}}
\end{picture}
\\
a)&&b)
\end{tabular}
\end{center}
\caption{Four-star configuration in four-dimensional Hilbert space
a) Greechie diagram representing atoms by points, and  contexts by maximal smooth, unbroken curves.
b) Dual Tkadlec diagram representing contexts by filled points, and interconnected contexts by lines.
\label{2006-omni-nstar} }
\end{figure}
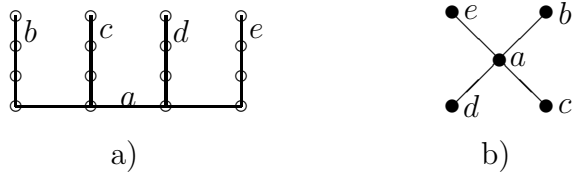

Tkadlec diagrams suggest the most compact representation of a context
in terms of a single maximal operator.
Note that, for the $n$-dimensional Hilbert space, an $n$-star configuration represents
$n$ different contexts joined in $n$ different atoms of the center context; see Fig.~\ref{2006-omni-nstar}.

\subsection{The ``one--zero'' rule}

\begin{figure*}%[p]
\begin{center}
\begin{tabular}{ccc}
%TeXCAD Picture [1.pic]. Options:
%\grade{\on}
%\emlines{\off}
%\epic{\off}
%\beziermacro{\on}
%\reduce{\on}
%\snapping{\off}
%\quality{8.00}
%\graddiff{0.01}
%\snapasp{1}
%\zoom{5.6569}
\unitlength .5mm % = 1.42pt
\linethickness{0.8pt}
\ifx\plotpoint\undefined\newsavebox{\plotpoint}\fi % GNUPLOT compatibility
\begin{picture}(120.92,114.73)(0,0)
%\emline(86.57,102.14)(111.57,58.64)
\multiput(86.57,102.14)(.11961722,-.20813397){209}{\line(0,-1){.20813397}}
%\end
%\emline(86.57,15.14)(111.57,58.64)
\multiput(86.57,15.14)(.11961722,.20813397){209}{\line(0,1){.20813397}}
%\end
%\emline(36.65,102.14)(11.65,58.64)
\multiput(36.65,102.14)(-.11961722,-.20813397){209}{\line(0,-1){.20813397}}
%\end
%\emline(36.65,15.14)(11.65,58.64)
\multiput(36.65,15.14)(-.11961722,.20813397){209}{\line(0,1){.20813397}}
%\end
\put(86.57,101.89){\line(-1,0){50}}
\put(86.57,15.39){\line(-1,0){50}}
\put(86.46,101.94){\circle{4}}
\put(86.46,15.34){\circle{4}}
\put(111.39,58.63){\circle{4}}
\put(11.74,58.63){\circle{4}}
\put(61.77,58.63){\circle{4}}
\put(36.52,101.94){\circle{4}}
\put(61.62,101.94){\circle{4}}
\put(61.62,15.44){\circle{4}}
\put(97.68,82.85){\circle{4}}
\put(25.71,82.85){\circle{4}}
\put(98.74,36.35){\circle{4}}
\put(24.65,36.35){\circle{4}}
\put(36.52,15.34){\circle{4}}
\put(61.69,101.82){\line(0,-1){86.27}}
\put(30.41,2.65){\makebox(0,0)[cc]{$A$}}
\put(61.87,2.3){\makebox(0,0)[cc]{$B$}}
\put(91.93,2.48){\makebox(0,0)[cc]{$C$}}
\put(110.84,30.94){\makebox(0,0)[cc]{$D$}}
\put(120.92,57.98){\makebox(0,0)[lc]{$E$}}
\put(108.41,88.92){\makebox(0,0)[cc]{$F$}}
\put(91.93,114.2){\makebox(0,0)[cc]{$G$}}
\put(61.7,114.73){\makebox(0,0)[cc]{$H$}}
\put(30.41,114.02){\makebox(0,0)[cc]{$I$}}
\put(13.56,87.86){\makebox(0,0)[cc]{$J$}}
\put(1.77,57.98){\makebox(0,0)[rc]{$ K$}}
\put(14.67,30.05){\makebox(0,0)[rc]{$L$}}
\put(67.88,55.51){\makebox(0,0)[cc]{$M$}}
\put(71.34,9.19){\makebox(0,0)[cc]{$a$}}
\put(107.91,40.35){\makebox(0,0)[cc]{$b$}}
\put(98.53,95.32){\makebox(0,0)[cc]{$c$}}
\put(54.46,108.01){\makebox(0,0)[cc]{$d$}}
\put(15.03,78.14){\makebox(0,0)[cc]{$e$}}
\put(21.56,27.06){\makebox(0,0)[cc]{$f$}}
\put(67.88,75.51){\makebox(0,0)[cc]{$g$}}
\end{picture}
&
$\qquad
$\qquad
$\qquad
$\qquad
&

%TeXCAD Picture [1.pic]. Options:
%\grade{\on}
%\emlines{\off}
%\epic{\off}
%\beziermacro{\on}
%\reduce{\on}
%\snapping{\off}
%\quality{8.00}
%\graddiff{0.01}
%\snapasp{1}
%\zoom{5.6569}
\unitlength .5mm % = 1.42pt
\linethickness{0.8pt}
\ifx\plotpoint\undefined\newsavebox{\plotpoint}\fi % GNUPLOT compatibility
\begin{picture}(120.92,114.2)(0,0)
%\emline(86.57,102.14)(111.57,58.64)
\multiput(86.57,102.14)(.11961722,-.20813397){209}{\line(0,-1){.20813397}}
%\end
%\emline(86.57,15.14)(111.57,58.64)
\multiput(86.57,15.14)(.11961722,.20813397){209}{\line(0,1){.20813397}}
%\end
%\emline(36.65,102.14)(11.65,58.64)
\multiput(36.65,102.14)(-.11961722,-.20813397){209}{\line(0,-1){.20813397}}
%\end
%\emline(36.65,15.14)(11.65,58.64)
\multiput(36.65,15.14)(-.11961722,.20813397){209}{\line(0,1){.20813397}}
%\end
\put(86.57,101.89){\line(-1,0){50}}
\put(86.57,15.39){\line(-1,0){50}}
\put(86.46,101.94){\circle*{4}}
\put(86.46,15.34){\circle*{4}}
\put(111.39,58.63){\circle*{4}}
\put(11.74,58.63){\circle*{4}}
\put(36.52,101.94){\circle*{4}}
\put(36.52,15.34){\circle*{4}}
\put(61.77,58.63){\circle*{4}}
%\emline(86.44,102)(36.59,15.56)
\multiput(86.44,102)(-.119831731,-.207788462){416}{\line(0,-1){.207788462}}
%\end
%-
\put(30.41,2.65){\makebox(0,0)[cc]{$a$}}
\put(91.93,2.48){\makebox(0,0)[cc]{$b$}}
\put(120.92,57.98){\makebox(0,0)[lc]{$c$}}
\put(91.93,114.2){\makebox(0,0)[cc]{$d$}}
\put(30.41,114.02){\makebox(0,0)[cc]{$e$}}
\put(1.77,57.98){\makebox(0,0)[rc]{$f$}}
\put(67.88,55.51){\makebox(0,0)[cc]{$g$}}
\end{picture}
\\
a)&&b)
\end{tabular}
\end{center}
\caption{Configuration of observables in three-dimensional Hilbert space implying
that whenever $K$ is true, $E$ must be false.
The seven interconnected contexts
$a=\{A,B,C\}$,
$b=\{C,D,E\}$,
$c=\{E,F,G\}$,
$d=\{G,H,I\}$,
$e=\{I,J,K\}$,
$f=\{K,L,A\}$,
$g=\{B,H,M\}$,
consist of the 13 projectors associated with the one dimensional subspaces spanned by
\cite{svozil-tkadlec}
$ A= ( 1,\sqrt{2},-1)      $,
$ B= ( 1,0,1)   $,
$ C= ( -1,\sqrt{2},1)    $,
$ D= ( -1,\sqrt{2},-3)$,
$  E=( \sqrt{2},1,0) $,
$  F=( 1,-\sqrt{2},-3)            $,
$  G=( -1,\sqrt{2},-1)           $,
$  H=( 1,0,-1)    $,
$  I=( 1,\sqrt{2},1)   $,
$ J= ( 1,\sqrt{2},-3)     $,
$ K=( \sqrt{2},-1,0)    $,
$ L=( 1,\sqrt{2},3)     $,
$ M=(0,1,0)    $.
%
%   kp[a1_, a2_, a3_, b1_, b2_, b3_] =   {a2  b3 - a3 b2, a3 b1 - a1 b3, a1 b2 - a2 b1}
%
a) Greechie diagram representing atoms by points, and  contexts by maximal smooth, unbroken curves.
Only a single observable per context can be true.
Noncontextuality requests that link observables in different contexts
are either true or false in all of these context.
Then, whenever $K$ is true, $E$ cannot be true, since then at least one of the two contexts $a$ and $d$
would contain only outcomes which do not occur.
b) Dual Tkadlec diagram representing contexts by filled points, and interconnected contexts by lines.
\label{2006-omni-oneimplieszero} }
\end{figure*}
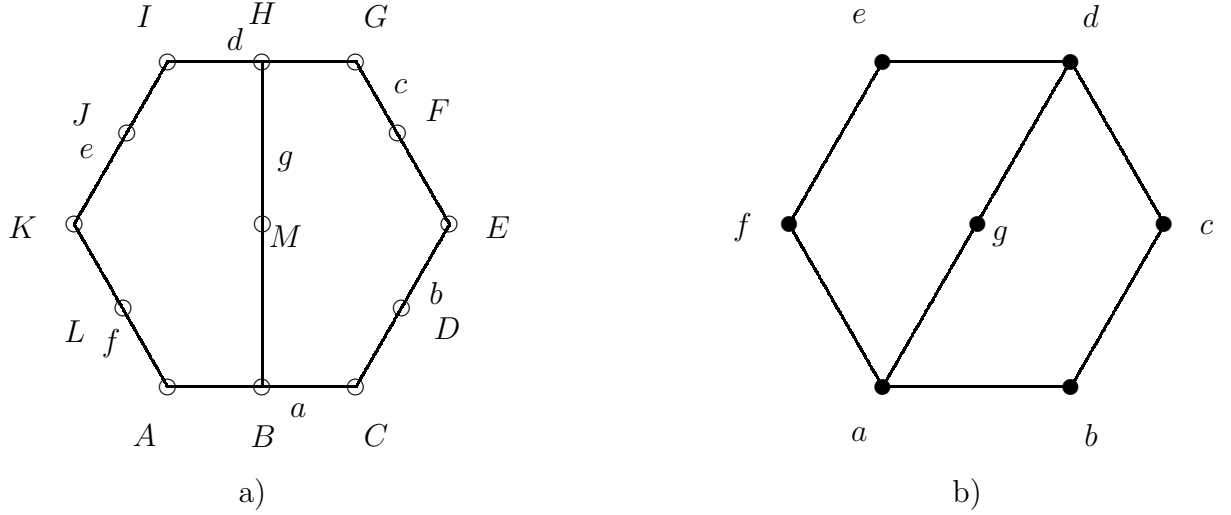

For the sake of presentation of such properties,
consider the proof that, for the observables depicted in Fig.~\ref{2006-omni-oneimplieszero},
the occurrence of an outcome corresponding to $K$ (abbreviated by ``$K$ occurs'')
implies that  $E$
cannot occur. This property,
which has been already exploited by KS
\cite[$\Gamma_1$]{kochen1}, will be called the ``one-zero rule.''

\subsection{The ``one--one/zero--zero'' rule}

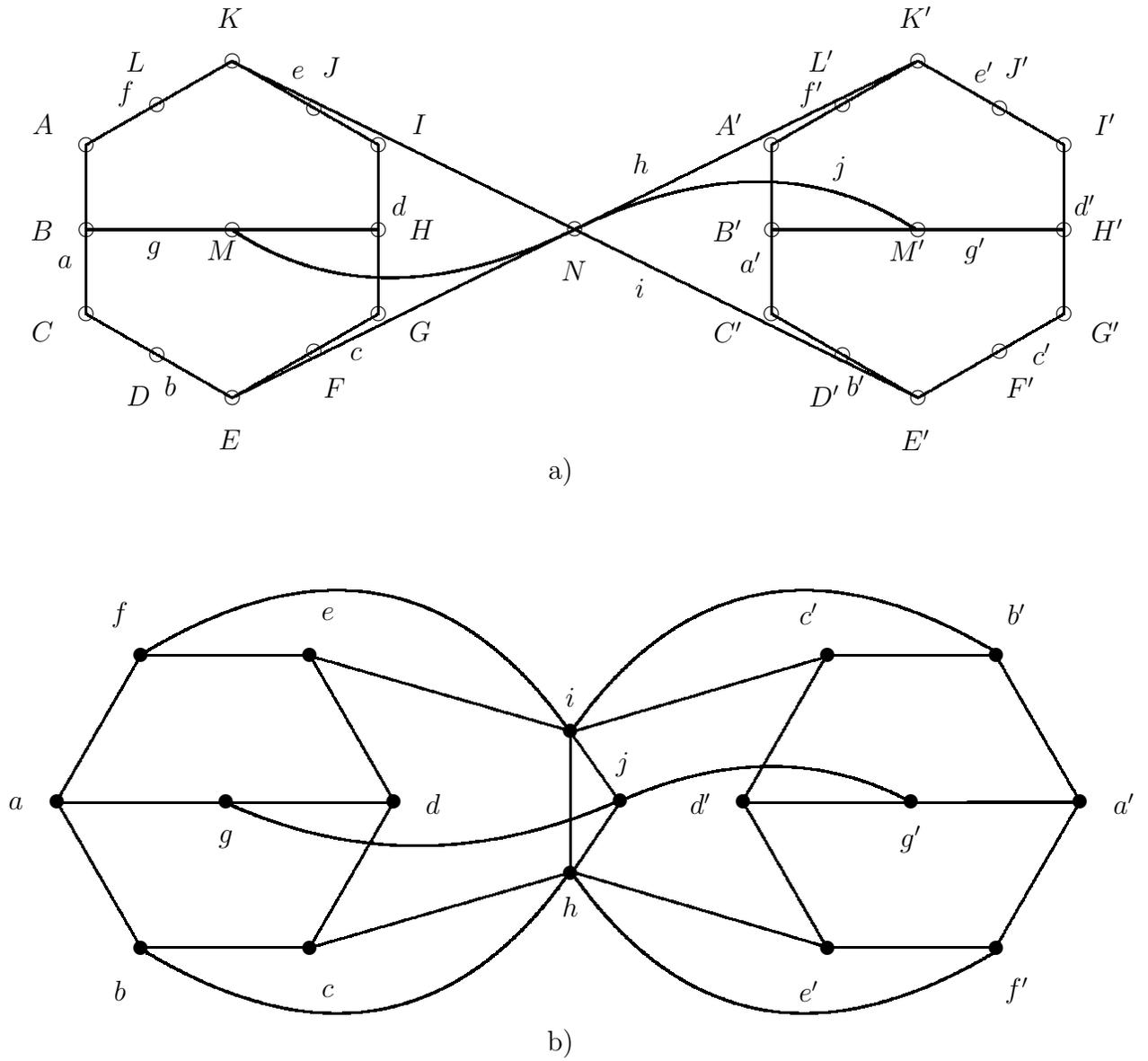
\begin{figure*}%[p]
\begin{center}
\begin{tabular}{c}

%TeXCAD Picture [1.pic]. Options:
%\grade{\on}
%\emlines{\off}
%\epic{\off}
%\beziermacro{\on}
%\reduce{\on}
%\snapping{\off}
%\quality{8.00}
%\graddiff{0.01}
%\snapasp{1}
%\zoom{4.0000}
\unitlength .5mm % = 1.42pt
\linethickness{0.8pt}
\ifx\plotpoint\undefined\newsavebox{\plotpoint}\fi % GNUPLOT compatibility
\begin{picture}(320.85,118.44)(0,0)
%\emline(105.32,33.64)(61.82,8.64)
\multiput(105.32,33.64)(-.20813397,-.11961722){209}{\line(-1,0){.20813397}}
%\end
%\emline(308.26,33.64)(264.76,8.64)
\multiput(308.26,33.64)(-.20813397,-.11961722){209}{\line(-1,0){.20813397}}
%\end
%\emline(18.32,33.64)(61.82,8.64)
\multiput(18.32,33.64)(.20813397,-.11961722){209}{\line(1,0){.20813397}}
%\end
%\emline(221.26,33.64)(264.76,8.64)
\multiput(221.26,33.64)(.20813397,-.11961722){209}{\line(1,0){.20813397}}
%\end
%\emline(105.32,83.56)(61.82,108.56)
\multiput(105.32,83.56)(-.20813397,.11961722){209}{\line(-1,0){.20813397}}
%\end
%\emline(308.26,83.56)(264.76,108.56)
\multiput(308.26,83.56)(-.20813397,.11961722){209}{\line(-1,0){.20813397}}
%\end
%\emline(18.32,83.56)(61.82,108.56)
\multiput(18.32,83.56)(.20813397,.11961722){209}{\line(1,0){.20813397}}
%\end
%\emline(221.26,83.56)(264.76,108.56)
\multiput(221.26,83.56)(.20813397,.11961722){209}{\line(1,0){.20813397}}
%\end
\put(105.07,33.64){\line(0,1){50}}
\put(308.01,33.64){\line(0,1){50}}
\put(18.57,33.64){\line(0,1){50}}
\put(221.51,33.64){\line(0,1){50}}
\put(105.12,33.75){\circle{4}}
\put(308.06,33.75){\circle{4}}
\put(18.52,33.75){\circle{4}}
\put(221.46,33.75){\circle{4}}
\put(61.81,8.82){\circle{4}}
\put(264.75,8.82){\circle{4}}
\put(61.81,108.47){\circle{4}}
\put(264.75,108.47){\circle{4}}
\put(61.81,58.44){\circle{4}}
\put(264.75,58.44){\circle{4}}
\put(105.12,83.69){\circle{4}}
\put(308.06,83.69){\circle{4}}
\put(105.12,58.59){\circle{4}}
\put(308.06,58.59){\circle{4}}
\put(18.62,58.59){\circle{4}}
\put(221.56,58.59){\circle{4}}
\put(86.03,22.53){\circle{4}}
\put(288.97,22.53){\circle{4}}
\put(86.03,94.5){\circle{4}}
\put(288.97,94.5){\circle{4}}
\put(39.53,21.47){\circle{4}}
\put(242.47,21.47){\circle{4}}
\put(39.53,95.56){\circle{4}}
\put(242.47,95.56){\circle{4}}
\put(18.52,83.69){\circle{4}}
\put(221.46,83.69){\circle{4}}
\put(163.21,58.69){\circle{4}}
\put(105,58.52){\line(-1,0){86.27}}
\put(307.94,58.52){\line(-1,0){86.27}}
\put(5.83,89.8){\makebox(0,0)[]{$A$}}
\put(208.77,89.8){\makebox(0,0)[]{$A'$}}
\put(5.48,58.34){\makebox(0,0)[]{$B$}}
\put(208.42,58.34){\makebox(0,0)[]{$B'$}}
\put(5.66,28.28){\makebox(0,0)[]{$C$}}
\put(208.6,28.28){\makebox(0,0)[]{$C'$}}
\put(34.12,9.37){\makebox(0,0)[]{$D$}}
\put(237.06,9.37){\makebox(0,0)[]{$D'$}}
\put(61.16,-.71){\makebox(0,0)[t]{$E$}}
\put(264.1,-.71){\makebox(0,0)[t]{$E'$}}
\put(92.1,11.8){\makebox(0,0)[]{$F$}}
\put(295.04,11.8){\makebox(0,0)[]{$F'$}}
\put(117.38,28.28){\makebox(0,0)[]{$G$}}
\put(320.32,28.28){\makebox(0,0)[]{$G'$}}
\put(117.91,58.51){\makebox(0,0)[]{$H$}}
\put(320.85,58.51){\makebox(0,0)[]{$H'$}}
\put(117.2,89.8){\makebox(0,0)[]{$I$}}
\put(320.14,89.8){\makebox(0,0)[]{$I'$}}
\put(91.04,106.65){\makebox(0,0)[]{$J$}}
\put(293.98,106.65){\makebox(0,0)[]{$J'$}}
\put(61.16,118.44){\makebox(0,0)[b]{$ K$}}
\put(264.1,118.44){\makebox(0,0)[b]{$ K'$}}
\put(33.23,105.54){\makebox(0,0)[b]{$L$}}
\put(236.17,105.54){\makebox(0,0)[b]{$L'$}}
\put(58.69,52.33){\makebox(0,0)[]{$M$}}
\put(261.63,52.33){\makebox(0,0)[]{$M'$}}
\put(12.37,48.87){\makebox(0,0)[]{$a$}}
\put(215.31,48.87){\makebox(0,0)[]{$a'$}}
\put(43.53,12.3){\makebox(0,0)[]{$b$}}
\put(246.47,12.3){\makebox(0,0)[]{$b'$}}
\put(98.5,21.68){\makebox(0,0)[]{$c$}}
\put(301.44,21.68){\makebox(0,0)[]{$c'$}}
\put(111.19,65.75){\makebox(0,0)[]{$d$}}
\put(314.13,65.75){\makebox(0,0)[]{$d'$}}
\put(81.32,105.18){\makebox(0,0)[]{$e$}}
\put(284.26,105.18){\makebox(0,0)[]{$e'$}}
\put(30.24,98.65){\makebox(0,0)[]{$f$}}
\put(233.18,98.65){\makebox(0,0)[]{$f'$}}
\put(38.69,52.33){\makebox(0,0)[]{$g$}}
\put(281.63,52.33){\makebox(0,0)[]{$g'$}}
%\emline(61.75,8.75)(264.5,108.5)
\multiput(61.75,8.75)(.243687933,.1198908573){832}{\line(1,0){.243687933}}
%\end
%\emline(61.75,108.25)(264.75,8.75)
\multiput(61.75,108.25)(.2445763352,-.1198785485){830}{\line(1,0){.2445763352}}
%\end
\put(163,46.25){\makebox(0,0)[cc]{$N$}}
\put(182.75,78.25){\makebox(0,0)[cc]{$h$}}
\put(182.5,41.25){\makebox(0,0)[cc]{$i$}}
\qbezier(61.75,58.5)(104.62,30)(163,58.5)
\qbezier(264.25,58.5)(221.37,87)(163,58.5)
\put(241.75,76.75){\makebox(0,0)[cc]{$j$}}
\end{picture}
\\

a)
\\

%TeXCAD Picture [1.pic]. Options:
%\grade{\on}
%\emlines{\off}
%\epic{\off}
%\beziermacro{\on}
%\reduce{\on}
%\snapping{\off}
%\quality{8.000}
%\graddiff{0.010}
%\snapasp{1}
%\zoom{4.0000}
\unitlength .5mm % = 1.423pt
\linethickness{0.8pt}
\ifx\plotpoint\undefined\newsavebox{\plotpoint}\fi % GNUPLOT compatibility
\begin{picture}(324.48,149.57)(0,0)
%\emline(86.57,102.14)(111.57,58.64)
\multiput(86.57,102.14)(.119617225,-.208133971){209}{\line(0,-1){.208133971}}
%\end
%\emline(239.68,102.14)(214.68,58.64)
\multiput(239.68,102.14)(-.119617225,-.208133971){209}{\line(0,-1){.208133971}}
%\end
%\emline(86.57,15.14)(111.57,58.64)
\multiput(86.57,15.14)(.119617225,.208133971){209}{\line(0,1){.208133971}}
%\end
%\emline(239.68,15.14)(214.68,58.64)
\multiput(239.68,15.14)(-.119617225,.208133971){209}{\line(0,1){.208133971}}
%\end
%\emline(36.65,102.14)(11.65,58.64)
\multiput(36.65,102.14)(-.119617225,-.208133971){209}{\line(0,-1){.208133971}}
%\end
%\emline(289.6,102.14)(314.6,58.64)
\multiput(289.6,102.14)(.119617225,-.208133971){209}{\line(0,-1){.208133971}}
%\end
%\emline(36.65,15.14)(11.65,58.64)
\multiput(36.65,15.14)(-.119617225,.208133971){209}{\line(0,1){.208133971}}
%\end
%\emline(289.6,15.14)(314.6,58.64)
\multiput(289.6,15.14)(.119617225,.208133971){209}{\line(0,1){.208133971}}
%\end
\put(86.57,101.89){\line(-1,0){50}}
\put(239.68,101.89){\line(1,0){50}}
\put(86.57,15.39){\line(-1,0){50}}
\put(239.68,15.39){\line(1,0){50}}
\put(86.46,101.94){\circle*{4}}
\put(163.71,79.44){\circle*{4}}
\put(178.46,58.94){\circle*{4}}
\put(163.71,37.56){\circle*{4}}
\put(239.79,101.94){\circle*{4}}
\put(86.46,15.34){\circle*{4}}
\put(239.79,15.34){\circle*{4}}
\put(111.39,58.63){\circle*{4}}
\put(214.86,58.63){\circle*{4}}
\put(11.74,58.63){\circle*{4}}
\put(314.51,58.63){\circle*{4}}
\put(36.52,101.94){\circle*{4}}
\put(289.73,101.94){\circle*{4}}
\put(36.52,15.34){\circle*{4}}
\put(289.73,15.34){\circle*{4}}
\put(61.77,58.63){\circle*{4}}
\put(264.48,58.63){\circle*{4}}
%-
%-
\put(30.41,2.65){\makebox(0,0)[cc]{$b$}}
\put(295.84,2.65){\makebox(0,0)[]{$f'$}}
\put(91.93,2.48){\makebox(0,0)[cc]{$c$}}
\put(234.32,2.48){\makebox(0,0)[]{$e'$}}
\put(120.92,57.98){\makebox(0,0)[lc]{$d$}}
\put(205.33,57.98){\makebox(0,0)[r]{$d'$}}
\put(91.93,114.2){\makebox(0,0)[cc]{$e$}}
\put(234.32,114.2){\makebox(0,0)[]{$c'$}}
\put(30.41,114.02){\makebox(0,0)[cc]{$f$}}
\put(295.84,114.02){\makebox(0,0)[]{$b'$}}
\put(1.77,57.98){\makebox(0,0)[rc]{$a$}}
\put(324.48,57.98){\makebox(0,0)[l]{$a'$}}
\put(61.77,47.51){\makebox(0,0)[cc]{$g$}}
\put(264.48,47.51){\makebox(0,0)[]{$g'$}}
\put(163.71,27.56){\makebox(0,0)[]{$h$}}
\put(163.71,89.44){\makebox(0,0)[]{$i$}}
\put(163.75,79.25){\line(0,-1){41.5}}
%\emline(86.8,15.38)(163.87,37.65)
\multiput(86.8,15.38)(.414354839,.119731183){186}{\line(1,0){.414354839}}
%\end
%\emline(86.8,101.62)(163.87,79.35)
\multiput(86.8,101.62)(.414354839,-.119731183){186}{\line(1,0){.414354839}}
%\end
%\emline(163.76,37.84)(239.87,15.35)
\multiput(163.76,37.84)(.404840426,-.11962766){188}{\line(1,0){.404840426}}
%\end
%\emline(163.76,79.16)(239.87,101.65)
\multiput(163.76,79.16)(.404840426,.11962766){188}{\line(1,0){.404840426}}
%\end
\qbezier(36.37,101.96)(114.36,149.57)(163.76,79.04)
\qbezier(36.37,15.04)(114.36,-32.57)(163.76,37.96)
\qbezier(291.16,101.96)(213.17,149.57)(163.76,79.04)
\qbezier(291.16,15.04)(213.17,-32.57)(163.76,37.96)
\put(12,58.25){\line(1,0){99.25}}
%\emline(215,58.25)(314.5,58.5)
\multiput(215,58.25)(33.16667,.08333){3}{\line(1,0){33.16667}}
%\end
%\emline(163.75,79.25)(178.5,58.5)
\multiput(163.75,79.25)(.119918699,-.168699187){123}{\line(0,-1){.168699187}}
%\end
%\emline(178.5,58.5)(164,37.25)
\multiput(178.5,58.5)(-.119834711,-.175619835){121}{\line(0,-1){.175619835}}
%\end
\qbezier(61.75,58.25)(118.375,32.375)(178.5,59)
\qbezier(178.5,58.75)(225.5,79.375)(264.5,58.5)
\put(179.25,69.75){\makebox(0,0)[cc]{$j$}}
\end{picture}
\\
b)
\end{tabular}
\end{center}
\caption{Configuration of observables implying that the occurrences of $K$ and $K'$ coincide.
a) Greechie diagram representing atoms by points, and  contexts by maximal smooth, unbroken curves.
The coordinates of the ``primed'' points $A'$--$M'$ are obtained by interchanging the first and the
second components of the unprimed coordinates $A$--$M$ enumerated in Fig.~\ref{2006-omni-oneimplieszero};
and $N=(0,0,1)$.
The two contexts $h$ and $i$ linking the primed with the unprimed observables allow the following argument:
Whenever $K$ occurs, then by the one-zero rule $E$ cannot occur;
moreover $N$ cannot occur, hence $K'$ must occur.
Conversely, by symmetry whenever $K'$ occurs, $K$ must occur.
b) Dual Tkadlec diagram representing contexts by filled points, and interconnected contexts by lines.
\label{2006-omni-oneimpliesone} }
\end{figure*}

For another example, consider
two collections of observables as above, which are combined by ``gluing'' them together in two contexts.
The geometry based upon the $\Gamma_3$-configuration in KS \cite{kochen1} is depicted in Fig.~\ref{2006-omni-oneimpliesone}.
In this case one obtains the ``one-one'' and ``zero-zero rules,''
stating that  $K$ occurs if and only if $K'$
occurs.

For a quantum falsification of the one-zero and the one-one/zero-zero  rules
it suffices to record a single pair of outcomes which does not obey these classical predictions.
This can for instance been demonstrated in an EPR-type setup of two spin one particles
in a singlet state $$\frac{1}{{\sqrt{3}}}\big(-|0,0\rangle+|-1,1\rangle+|1,-1\rangle\big),$$
and observables corresponding to $E$ and $K$, or to $K$ and $K'$.
Generalized beam splitters  are possible realizations \cite{rzbb,zukowski-97,svozil-2004-analog}.
This adds to the evidence accumulated already by Bell, KS and GHZ, that we are not living in a classical world.

\subsection{The absence of two--valued states}

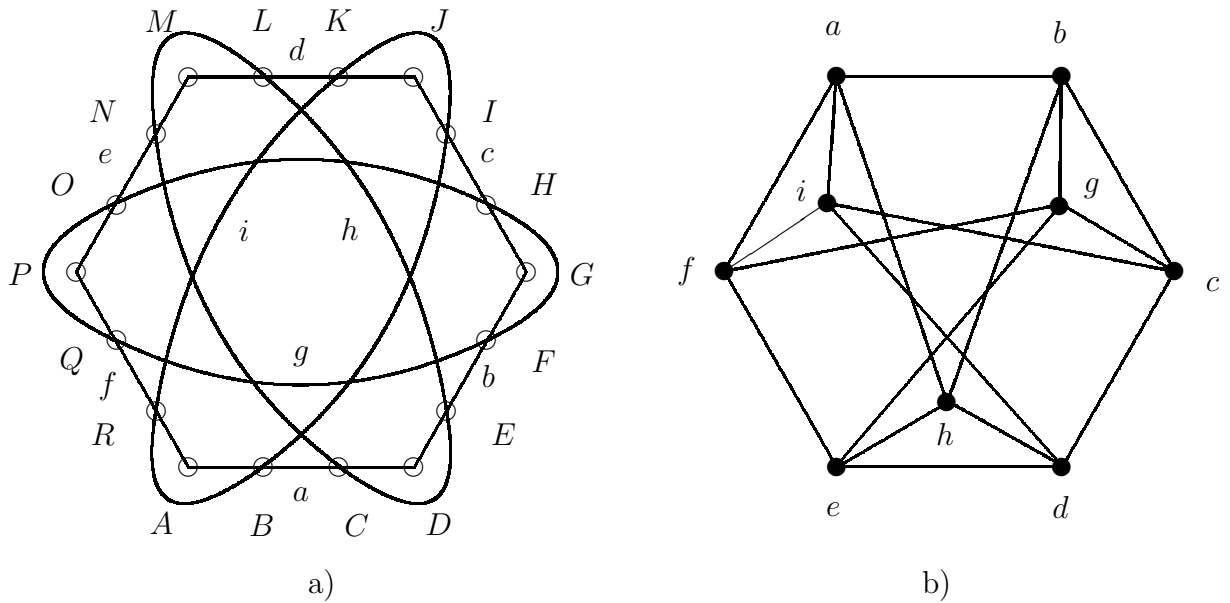
\begin{figure*}
\begin{center}
\begin{tabular}{cc}
%TeXCAD Picture [1.pic]. Options:
%\grade{\on}
%\emlines{\off}
%\epic{\off}
%\beziermacro{\on}
%\reduce{\on}
%\snapping{\off}
%\quality{8.000}
%\graddiff{0.010}
%\snapasp{1}
%\zoom{5.6569}
\unitlength .6mm % = 1.423pt
\linethickness{0.8pt}
\ifx\plotpoint\undefined\newsavebox{\plotpoint}\fi % GNUPLOT compatibility
\begin{picture}(134.09,125.99)(0,0)
%\emline(86.39,101.96)(111.39,58.46)
\multiput(86.39,101.96)(.119617225,-.208133971){209}{\line(0,-1){.208133971}}
%\end
%\emline(86.39,14.96)(111.39,58.46)
\multiput(86.39,14.96)(.119617225,.208133971){209}{\line(0,1){.208133971}}
%\end
%\emline(36.47,101.96)(11.47,58.46)
\multiput(36.47,101.96)(-.119617225,-.208133971){209}{\line(0,-1){.208133971}}
%\end
%\emline(36.47,14.96)(11.47,58.46)
\multiput(36.47,14.96)(-.119617225,.208133971){209}{\line(0,1){.208133971}}
%\end
\put(86.39,101.71){\line(-1,0){50}}
\put(86.39,15.21){\line(-1,0){50}}
\put(86.28,101.76){\circle{4}}
\put(86.28,15.16){\circle{4}}
\put(93.53,89.21){\circle{4}}
\put(93.53,27.71){\circle{4}}
\put(29.24,89.21){\circle{4}}
\put(29.24,27.71){\circle{4}}
\put(102.37,73.47){\circle{4}}
\put(102.37,43.44){\circle{4}}
\put(20.4,73.47){\circle{4}}
\put(20.4,43.44){\circle{4}}
\put(111.21,58.45){\circle{4}}
\put(11.56,58.45){\circle{4}}
\put(36.34,101.76){\circle{4}}
\put(36.34,15.16){\circle{4}}
\put(52.99,101.76){\circle{4}}
\put(52.99,15.16){\circle{4}}
\put(69.68,101.76){\circle{4}}
\put(69.68,15.16){\circle{4}}
\qbezier(29.2,27.73)(23.55,-5.86)(52.99,15.24)
\qbezier(93.57,27.73)(99.22,-5.86)(69.78,15.24)
\qbezier(29.2,27.88)(36.93,75)(69.63,101.91)
\qbezier(93.57,27.88)(85.84,75)(53.13,101.91)
\qbezier(52.69,15.24)(87.47,40.96)(93.72,89.27)
\qbezier(70.08,15.24)(35.3,40.96)(29.05,89.27)
\qbezier(93.72,89.27)(98.4,125.99)(69.49,102.06)
\qbezier(29.05,89.27)(24.37,125.99)(53.28,102.06)
\qbezier(20.15,73.72)(-11.67,58.52)(20.15,43.31)
\qbezier(20.33,73.72)(61.34,93.16)(102.36,73.72)
\qbezier(102.36,73.72)(134.09,58.52)(102.53,43.31)
\qbezier(102.53,43.31)(60.99,23.43)(20.15,43.49)
\put(30.41,114.02){\makebox(0,0)[cc]{$M$}}
\put(30.41,2.65){\makebox(0,0)[cc]{$A$}}
\put(52.68,114.38){\makebox(0,0)[cc]{$L$}}
\put(52.68,2.3){\makebox(0,0)[cc]{$B$}}
\put(91.93,114.2){\makebox(0,0)[cc]{$J$}}
\put(91.93,2.48){\makebox(0,0)[cc]{$D$}}
\put(69.65,114.38){\makebox(0,0)[cc]{$K$}}
\put(73.65,2.3){\makebox(0,0)[cc]{$C$}}
\put(103.24,94.22){\makebox(0,0)[cc]{$I$}}
\put(17.45,94.22){\makebox(0,0)[cc]{$ N$}}
\put(106.24,22.45){\makebox(0,0)[cc]{$E$}}
\put(17.45,22.45){\makebox(0,0)[cc]{$ R$}}
\put(115.13,77.96){\makebox(0,0)[cc]{$H$}}
\put(8.55,77.96){\makebox(0,0)[cc]{$ O$}}
\put(115.13,38.72){\makebox(0,0)[cc]{$F$}}
\put(10.55,38.72){\makebox(0,0)[cc]{$ Q$}}
\put(120.92,57.98){\makebox(0,0)[l]{$ G$}}
\put(1.77,57.98){\makebox(0,0)[rc]{$  P$}}
\put(61.341,9.192){\makebox(0,0)[cc]{$a$}}
\put(102.883,35.355){\makebox(0,0)[cc]{$b$}}
\put(102.53,84.322){\makebox(0,0)[cc]{$c$}}
\put(60.457,108.01){\makebox(0,0)[cc]{$d$}}
\put(18.031,84.145){\makebox(0,0)[cc]{$e$}}
\put(18.561,33.057){\makebox(0,0)[cc]{$f$}}
\put(61.341,39.774){\makebox(0,0)[cc]{$g$}}
\put(72.124,67.882){\makebox(0,0)[cc]{$h$}}
\put(48.79,67.705){\makebox(0,0)[cc]{$i$}}
\end{picture}
&
%TeXCAD Picture [1.pic]. Options:
%\grade{\on}
%\emlines{\off}
%\epic{\off}
%\beziermacro{\on}
%\reduce{\on}
%\snapping{\off}
%\quality{8.000}
%\graddiff{0.010}
%\snapasp{1}
%\zoom{5.6569}
\unitlength .6mm % = 1.423pt
\linethickness{0.8pt}
\ifx\plotpoint\undefined\newsavebox{\plotpoint}\fi % GNUPLOT compatibility
\begin{picture}(119.854,112.606)(0,0)
%\emline(86.567,102.137)(111.567,58.637)
\multiput(86.567,102.137)(.119617225,-.208133971){209}{\line(0,-1){.208133971}}
%\end
%\emline(86.567,15.137)(111.567,58.637)
\multiput(86.567,15.137)(.119617225,.208133971){209}{\line(0,1){.208133971}}
%\end
%\emline(36.647,102.137)(11.647,58.637)
\multiput(36.647,102.137)(-.119617225,-.208133971){209}{\line(0,-1){.208133971}}
%\end
%\emline(36.647,15.137)(11.647,58.637)
\multiput(36.647,15.137)(-.119617225,.208133971){209}{\line(0,1){.208133971}}
%\end
\put(86.567,101.887){\line(-1,0){50}}
\put(86.567,15.387){\line(-1,0){50}}
\put(86.457,101.937){\circle*{4}}
\put(86.457,15.337){\circle*{4}}
\put(85.927,73.127){\circle*{4}}
\put(34.486,73.849){\circle*{4}}
\put(111.387,58.627){\circle*{4}}
\put(11.737,58.627){\circle*{4}}
\put(36.517,101.937){\circle*{4}}
\put(36.517,15.337){\circle*{4}}
\put(60.941,29.586){\circle*{4}}
%\emline(86.087,73.357)(86.447,101.997)
\multiput(86.087,73.357)(.12,9.54667){3}{\line(0,1){9.54667}}
%\end
%\emline(86.267,73.537)(111.187,58.687)
\multiput(86.267,73.537)(.200967742,-.119758065){124}{\line(1,0){.200967742}}
%\end
%\emline(86.087,73.357)(11.667,58.517)
\multiput(86.087,73.357)(-.60016129,-.119677419){124}{\line(-1,0){.60016129}}
%\end
%\emline(86.087,73.187)(36.237,15.207)
\multiput(86.087,73.187)(-.1198317308,-.139375){416}{\line(0,-1){.139375}}
%\end
%\emline(36.951,15.376)(61.341,29.696)
\multiput(36.951,15.376)(.20325,.119333333){120}{\line(1,0){.20325}}
%\end
%\emline(61.341,29.696)(86.801,15.376)
\multiput(61.341,29.696)(.212166667,-.119333333){120}{\line(1,0){.212166667}}
%\end
%\emline(60.991,29.696)(36.591,101.296)
\multiput(60.991,29.696)(-.119607843,.350980392){204}{\line(0,1){.350980392}}
%\end
%\emline(61.161,29.526)(86.801,101.646)
\multiput(61.161,29.526)(.119813084,.337009346){214}{\line(0,1){.337009346}}
%\end
\put(11.844,58.69){\line(3,2){22.804}}
%\emline(34.648,73.892)(36.593,101.823)
\multiput(34.648,73.892)(.1144118,1.643){17}{\line(0,1){1.643}}
%\end
%\emline(86.62,15.38)(34.472,73.716)
\multiput(86.62,15.38)(-.1198804598,.1341057471){435}{\line(0,1){.1341057471}}
%\end
%\emline(34.472,73.716)(111.369,58.513)
\multiput(34.472,73.716)(.605488189,-.119708661){127}{\line(1,0){.605488189}}
%\end
\put(35.885,5.834){\makebox(0,0)[cc]{$e$}}
\put(86.266,6.364){\makebox(0,0)[cc]{$d$}}
\put(119.854,55.861){\makebox(0,0)[cc]{$c$}}
\put(86.09,111.722){\makebox(0,0)[cc]{$b$}}
\put(35.885,112.606){\makebox(0,0)[cc]{$a$}}
\put(3.359,58.689){\makebox(0,0)[cc]{$f$}}
\put(60.634,22.45){\makebox(0,0)[cc]{$h$}}
\put(93.161,77.074){\makebox(0,0)[cc]{$g$}}
\put(28.814,76.544){\makebox(0,0)[cc]{$i$}}
\end{picture}
\\
a)&b)\\
\end{tabular}
\end{center}
\caption{Proof of the Kochen-Specker theorem \cite{cabello-96,cabello-99} in four-dimensional real vector space.
The nine tightly interconnected contexts
$a=\{A,B,C,D\}$,
$b=\{D,E,F,G\}$,
$c=\{G,H,I,J\}$,
$d=\{J,K,L,M\}$,
$e=\{M,N,O,P\}$,
$f=\{P,Q,R,A\}$,
$g=\{B,I,K,R\}$,
$h=\{C,E,L,N\}$,
$i=\{F,H,O,Q\}$
consist of the 18 projectors associated with the one dimensional subspaces spanned by
$ A=(0,0,1,-1)    $,
$ B=(1,-1,0,0)    $,
$ C=(1,1,-1,-1)   $,
$ D=(1,1,1,1)     $,
$  E=(1,-1,1,-1)  $,
$  F=(1,0,-1,0)   $,
$  G=(0,1,0,-1)   $,
$  H=(1,0,1,0)    $,
$  I=(1,1,-1,1)   $,
$ J=(-1,1,1,1)    $,
$ K=(1,1,1,-1)    $,
$ L=(1,0,0,1)     $,
$ M=(0,1,-1,0)    $,
$  N=(0,1,1,0)    $,
$  O=(0,0,0,1)    $,
$  P=(1,0,0,0)    $,
$  Q=(0,1,0,0)    $,
$  R=(0,0,1,1)    $.
a) Greechie diagram representing atoms by points, and  contexts by maximal smooth, unbroken curves.
b) Dual Tkadlec diagram representing contexts by filled points, and interconnected contexts are connected by lines.
(Duality means that points represent blocks and maximal smooth curves represent atoms.)
The nine contexts in four dimensional Hilbert space are interlinked in a four-star form;
hence every observable proposition occurs in exactly two contexts.
Thus, in an enumeration of the four observable propositions of each of the nine contexts,
there appears to be an {\em even} number of true propositions.
Yet, as there is an odd number of contexts,
there should be an {\em odd} number (actually nine) of true propositions.
\label{2007-miracles-ksc} }
\end{figure*}

The simplest known proof \cite{cabello-96,cabello-99} of KS is  in four-dimensional real Hilbert space
and requires nine intricately interwoven contexts --- every observable is in exactly two different contexts ---
 drawn in Fig.~\ref{2007-miracles-ksc}.
In order to appreciate the proofs (by contradiction), note that
\begin{itemize}

\item[(i)] the proofs require the assumption of counterfactuals;
i.e., of ``potential'' observables which, due to quantum complementarity,
are incompatible with the ``actual'' measurement context;
yet could have been
measured if the measurement apparatus were different.
These counterfactuals are  organized into groups of interconnected contexts which,
due to quantum complementarity, are incompatible
and therefore cannot be measured simultaneously; not even in Einstein-Podolsky-Rosen (EPR)
\cite{epr}
type setups
\cite{svozil-2006-uniquenessprinciple}.

\item[(ii)]   The proofs by contradiction have no direct experimental realizations.
As has already been pointedly stated by Robert Clifton
\cite{clifton},
``how can you measure a contradiction?''

\item[(iii)] So--called ``experimental tests'' inspired by Bell-type inequalities
\cite{aspect-81,aspect-82a,wjswz-98},
 KS
\cite{simon-2002,hasegawa:230401}
as well as GHZ
\cite{panbdwz}
measure the incompatible contexts which are considered  in the proofs one after another;
i.e., temporally sequentially, and not simultaneously.
Hence, different contexts can only be measured on different particles.
\end{itemize}

\section{Alternatives}

The following alternatives present some ways to cope with these findings:

\begin{itemize}

\item[(i)] abandonment of classical omniscience: in this view, it is wrong to assume that
all observables which could in principle (``potentially'') have been measured also co--exist,
irrespective of whether or not they have or even could have been actually measured.
Realism might still be assumed for a {\em single} context, in particular the one in which the system was prepared;

\item[(ii)]   abandonment of realism: in this view, it is wrong to assume that physical entities exist
even without being experienced by any finite mind.
Quite literary, with this assumption, the proofs of KS and similar decay into thin air because
there are no counterfactuals or unobserved physical observables
or inferred (rather than measured) elements of physical reality.

\item[(iii)] contextuality; i.e., the abandonment of context independence of measurement outcomes \cite{bell-66,hey-red,redhead}:
it is wrong to assume (cf. Ref.~\cite{bell-66}, Sec.~5) that the
result of an observation is independent
not only of the state of the system
but also of the complete disposition  of the apparatus.
Compare also Bohr's
remarks \cite{bohr-1949}
about {\em ``the impossibility of any sharp separation
between the behavior of atomic objects and the interaction with the measuring instruments which serve to define
the conditions under which the phenomena appear.''}

\end{itemize}

It should come as no surprise that realists
such as Bell favor contextuality rather than giving up realism or classical omniscience.
Nonetheless, to this date there does not exist a single experimental finding
to support contextuality, and, as pointed out above, contextuality is
only one of at least three possibilities to interpret  quantum probability theory.

The simplest configuration testing contextuality corresponds to an arrangement
of five observables $A,B,C,D,K$ with two comeasurable, mutually commuting, contexts
$\{A,B,C\}$
and
$\{A,D,K\}$
interconnected at $A$.
This propositional structure $L_{12}$ can be represented in three-dimensional Hilbert space
by two tripods with a single common leg.
Indeed, if contextuality is a physically meaningful principle
for the finite systems of observables employed in proofs of KS,
then  contextuality should  already be
detectable in this simple system of observables.
It would be a challenging task to realize the $L_{12}$ quantum logical structure experimentally
in an EPR-type setup, and falsify contextuality there.

Furthermore, in extension of the two-context configuration,
also systems of three interlinked contexts such as
$\{A,B,C\}$,
$\{A,D,K\}$
and
$\{K,L,M\}$ interconnected at $A$ and $K$
could be considered.
Note that too tightly interconnected systems such as
$\{A,B,C\}$,
$\{A,D,K\}$
and
$\{K,L,C\}$
have no representation in a 3-dimensional Hilbert space.
However, for a greater dimension than three, we can take,
e.g., $A=(1,0,0,0)$, $B=(0,1,0,0)$, $C=(0,0,1,0)$, $D=(0,1,1,0)$, $K=(0,0,0,1)$, $L=(1,1,0,0)$.

\section{Summary}

If one believes  in the physical existence of counterfactuals,
a lot of puzzling and mindboggling properties can be derived, bordering to mystery,
if not to  absurdity.
Take, for example the one-zero rule discussed above:
a noncontextual argument shows that certain outcomes are correlated.

Formally, this is due to the ``scarcity'' of two--valued states on the linear subspaces of Hilbert states.
Worse yet, by considering a larger, finite group of observables, it can be shown that, with the assumption
of noncontextuality, no such state exists.

Alas, it is not too difficult to derive ``mindboggling'' statements from absurdities.
Indeed, the  {\em principle of explosion} suggests that ``anything follows from a contradiction.''

It is not unreasonable to doubt the usefulness of
contextuality as a resolution of the imminent inconsistencies and complete contradictions originating in
the assumption of the physical (co-)existence of observables in different contexts.
Contextuality might not even be measurable in the simplest cases where
it could be falsified by simultaneous EPR-type measurements of two interlinked contexts.
A detailed discussion on realism versus empiricism and
the issues related to contextuality in EPR-type configurations
can also be found in Refs.~\cite{Muynck2001,Muynck2002};
see also Khrennikov's findings about couterfactuals in EPR-type setups~\cite{Khrennikov2008}.

It appears most natural to abandon the notion that not all classical observables
are quantum observables; that quantum omniscience is limited to a single context;
that a quantized system has only observable physical properties in the context in which it was prepared;
and that one should accept the obvious fact that one cannot squeeze information
from an ignorant system or agent.
If one tries nevertheless, then all one obtains are random, erratic outcomes.
Indeed, it is not totally unreasonable to speculate that contextuality is a ``red herring;''
that it appears to be one of the biggest and most popular
delusions in the foundations of the quantum (which is rich in mindboggling speculations),
devised by Bell and other realist to retain some form of classical realistic nonsensical omniscience.

\section*{Acknowledgements}
The author gratefully acknowledges the suggestions of two anonymous referees.

%\bibliography{svozil}
%\bibliographystyle{osa}
%\bibliographystyle{apsrev}
%\bibliographystyle{unsrt}
%\bibliographystyle{plain}

\end{document}